\documentclass[12pt]{spieman}  
\usepackage{amsmath,amsfonts,amssymb}
\usepackage{graphicx}
\usepackage{setspace}
\usepackage{tocloft}
\usepackage{orcidlink}
\usepackage{multirow}
\usepackage{multicol}
\usepackage{xspace}

\usepackage{lineno}
\newcommand{\daksha}{\textit{Daksha}}
\newcommand{\asat}{{\em AstroSat}}

\newcommand{\fermi}{\emph{Fermi}}

\newcommand{\polar}{\emph{POLAR}}
\newcommand{\integral}{\emph{INTEGRAL}}

\newcommand{\geant}{\textsc{GEANT4\xspace}}
\newcommand{\mcmdp}{MC-MDP\xspace}
\newcommand{\degr}{\ensuremath{^\circ}}

\newcommand{\ec}{\ensuremath{\mathrm{erg~cm}^{-2}}}

\newcommand{\citep}[1]{[\citen{#1}]}

\makeatletter
\let\jnl@style=\rm
\def\ref@jnl#1{{\jnl@style#1}}

\def\aj{\ref@jnl{AJ}}                   
\def\actaa{\ref@jnl{Acta Astron.}}      
\def\araa{\ref@jnl{ARA\&A}}             
\def\apj{\ref@jnl{ApJ}}                 
\def\apjl{\ref@jnl{ApJ}}                
\def\apjs{\ref@jnl{ApJS}}               
\def\ao{\ref@jnl{Appl.~Opt.}}           
\def\apss{\ref@jnl{Ap\&SS}}             
\def\aap{\ref@jnl{A\&A}}                
\def\aapr{\ref@jnl{A\&A~Rev.}}          
\def\aaps{\ref@jnl{A\&AS}}              
\def\azh{\ref@jnl{AZh}}                 
\def\baas{\ref@jnl{BAAS}}               
\def\bac{\ref@jnl{Bull. astr. Inst. Czechosl.}}
\def\caa{\ref@jnl{Chinese Astron. Astrophys.}}
\def\cjaa{\ref@jnl{Chinese J. Astron. Astrophys.}}
\def\icarus{\ref@jnl{Icarus}}           
\def\jcap{\ref@jnl{J. Cosmology Astropart. Phys.}}
\def\jrasc{\ref@jnl{JRASC}}             
\def\memras{\ref@jnl{MmRAS}}            
\def\mnras{\ref@jnl{MNRAS}}             
\def\na{\ref@jnl{New A}}                
\def\nar{\ref@jnl{New A Rev.}}          
\def\pra{\ref@jnl{Phys.~Rev.~A}}        
\def\prb{\ref@jnl{Phys.~Rev.~B}}        
\def\prc{\ref@jnl{Phys.~Rev.~C}}        
\def\prd{\ref@jnl{Phys.~Rev.~D}}        
\def\pre{\ref@jnl{Phys.~Rev.~E}}        
\def\prl{\ref@jnl{Phys.~Rev.~Lett.}}    
\def\pasa{\ref@jnl{PASA}}               
\def\pasp{\ref@jnl{PASP}}               
\def\pasj{\ref@jnl{PASJ}}               
\def\rmxaa{\ref@jnl{Rev. Mexicana Astron. Astrofis.}}%
\def\qjras{\ref@jnl{QJRAS}}             
\def\skytel{\ref@jnl{S\&T}}             
\def\solphys{\ref@jnl{Sol.~Phys.}}      
\def\sovast{\ref@jnl{Soviet~Ast.}}      
\def\ssr{\ref@jnl{Space~Sci.~Rev.}}     
\def\zap{\ref@jnl{ZAp}}                 
\def\nat{\ref@jnl{Nature}}              
\def\iaucirc{\ref@jnl{IAU~Circ.}}       
\def\aplett{\ref@jnl{Astrophys.~Lett.}} 
\def\apspr{\ref@jnl{Astrophys.~Space~Phys.~Res.}}
\def\bain{\ref@jnl{Bull.~Astron.~Inst.~Netherlands}} 
\def\fcp{\ref@jnl{Fund.~Cosmic~Phys.}}  
\def\gca{\ref@jnl{Geochim.~Cosmochim.~Acta}}   
\def\grl{\ref@jnl{Geophys.~Res.~Lett.}} 
\def\jcp{\ref@jnl{J.~Chem.~Phys.}}      
\def\jgr{\ref@jnl{J.~Geophys.~Res.}}    
\def\jqsrt{\ref@jnl{J.~Quant.~Spec.~Radiat.~Transf.}}
\def\memsai{\ref@jnl{Mem.~Soc.~Astron.~Italiana}}
\def\nphysa{\ref@jnl{Nucl.~Phys.~A}}   
\def\physrep{\ref@jnl{Phys.~Rep.}}   
\def\physscr{\ref@jnl{Phys.~Scr}}   
\def\planss{\ref@jnl{Planet.~Space~Sci.}}   
\def\procspie{\ref@jnl{Proc.~SPIE}}   

\makeatother

\title{Prospects of measuring Gamma-ray Burst Polarisation with the \daksha{} mission}

\author[a*]{Suman Bala~\orcidlink{0000-0002-6657-9022}}
\author[b*]{Sujay Mate~\orcidlink{0000-0001-5536-4635}}
\author[a]{Advait Mehla~\orcidlink{0000-0002-3155-6584}}
\author[a,c]{Parth Sastry~\orcidlink{0000-0001-9645-5453}}
\author[d]{N. P. S. Mithun~\orcidlink{0000-0003-3431-6110}}
\author[a,e]{Sourav Palit~\orcidlink{0000-0003-2932-3666}}
\author[a]{Mehul Vijay Chanda~\orcidlink{0000-0001-6343-1674}}
\author[a]{Divita Saraogi~\orcidlink{0000-0001-6332-1723}}
\author[d]{C. S. Vaishnava~\orcidlink{0000-0002-8096-5683}}
\author[a]{Gaurav Waratkar~\orcidlink{0000-0003-3630-9440}}
\author[a]{Varun Bhalerao~\orcidlink{0000-0002-6112-7609}}
\author[f]{Dipankar Bhattacharya~\orcidlink{0000-0003-3352-3142}}
\author[b]{Shriharsh Tendulkar~\orcidlink{0000-0003-2548-2926}}
\author[d]{Santosh Vadawale~\orcidlink{0000-0002-2050-0913}}

\affil[a]{Department of Physics, IIT Bombay, Powai, Mumbai, 400076, India.}
\affil[b]{Department of Astronomy and Astrophysics, Tata Institute of Fundamental Research, Mumbai, 400005, India.}
\affil[c]{Department of Physics, University of Massachusetts, 285 Old Westport Road, Dartmouth, 02747-2300, USA.}
\affil[d]{Physical Research Laboratory, Navrangpura, Ahmedabad, 380009, India.}
\affil[e]{Indian Centre for Space Physics, Mukundapur, Kolkata, 700099, India.}
\affil[f]{Ashoka University, Department of Physics, Sonepat, Haryana, 131029, India.}

\cftpagenumbersoff{figure}
\cftpagenumbersoff{table}
\begin{document}
\maketitle

\begin{abstract}
The proposed \daksha{} mission comprises of a pair of highly sensitive space telescopes for detecting and characterising high-energy transients such as electromagnetic counterparts of gravitational wave events and gamma-ray bursts (GRBs). Along with spectral and timing analysis, \daksha{} can also undertake polarisation studies of these transients, providing data crucial for understanding the source geometry and physical processes governing high-energy emission. Each \daksha{} satellite will have 340 pixelated Cadmium Zinc Telluride (CZT) detectors arranged in a quasi-hemispherical configuration without any field-of-view collimation (open detectors). These CZT detectors are good polarimeters in the energy range 100 – 400 keV, and their ability to measure polarisation has been successfully demonstrated by the Cadmium Zinc Telluride Imager (CZTI) onboard \asat. Here we demonstrate the hard X-ray polarisation measurement capabilities of \daksha{} and estimate the polarisation measurement sensitivity (in terms of the Minimum Detectable Polarisation: MDP) using extensive simulations. We find that \daksha{} will have MDP of~$30\%$ for a fluence threshold of $10^{-4}$~\ec~(in 10 -- 1000 keV). We estimate that with this sensitivity, if GRBs are highly polarised, \daksha{} can measure the polarisation of about five GRBs per year.
\end{abstract}


{\noindent \footnotesize\textbf{*}Suman Bala, \linkable{sumanbala2210@gmail.com}\\
\noindent \footnotesize\textbf{*}Sujay Mate, \linkable{sujay.mate@tifr.res.in}\\
\noindent \footnotesize \textbf{*} These authors contributed equally to this work.}

\begin{spacing}{1}   

\section{Introduction}\label{sec:intro}
Gamma-Ray Bursts~\cite[GRBs]{discovery} are the most energetic explosions in the Universe. The initial brief and intense gamma-ray flash originates close to the burst site and is known as the prompt emission. When burst ejecta collides with the external ambient medium, it produces the afterglow emission, which radiates in all wavelengths (radio to gamma-rays)~\citep{Rees1992,piran05,meszaros06}. GRBs are classified into two categories depending upon the duration of the prompt emission phase and they originate from two different progenitors. Short GRBs are produced by the merger of two compact objects, such as binary neutron stars (BNS) or a neutron star-black hole (NS-BH) pair~\citep{eichler89, narayan92, Abbott_170817A}, whereas long GRBs originate from the core collapse of massive stars~\citep{woosley93, iwamoto98, macfadyen99}. In both cases, the central engine is believed to be either a black hole~\citep{woosley93, iwamoto98,macfadyen99} or a hyper-massive magnetar~\citep{eichler89,narayan92}, that is thought to emit a relativistic jet giving rise to the GRB.

We have a broad understanding of GRBs, but many detailed questions regarding the nature of the central engine and the emission from the relativistic jet still remain unanswered~\citep{kumar15,zhang19}. There are competing models regarding the exact energy dissipation process, radiation mechanism, and radiation transfer processes in the prompt emission phase~\citep{Gill_etal_2021_review}. It is well established that these combined processes produce a non-thermal spectrum, often fitted with various phenomenological models like a power law, power law with an exponential cutoff, Band function,  etc. \citep{band1993batse,gruber2014fermi}. Most of the spectral analyses show similar fit statistics when fitted with different physical and empirical models \citep{iyyani15, Zhang2016} ; however, different physical models predict different degrees of polarization. Hence a statistical analysis of polarisation properties of the GRB prompt emission phase can be a very useful tool to constrain the models and answer questions regarding the emission mechanism and geometry of GRB jets \citep{toma08, covino16, mcconnell16, gill18, Gill_etal_2021_review}.

High-energy polarisation measurement is a very photon-hungry task. Due to the short-lived nature of the prompt emission phase, it has been extremely difficult to measure polarisation of GRBs. Polarisation measurements exits only for $\sim40$ GRBs out of $>5000$ GRBs detected so far. The majority of these detections have come only in the last few years with \polar~\citep{zhang19,Kole2020} and \asat/CZTI~\citep{chattopadhyay19,Chattopadhyay2022} leading the numbers and other missions like \emph{BATSE}~\citep{Willis2005}, \emph{GAP}~\citep{Yonetoku2012} and \integral/SPI~\citep{Gotz09,Gotz13} contributing to a few measurements. However, different missions have measured different levels of polarisation in the observed GRBs. For example, \emph{GAP} and \emph{INTEGRAL} have measured high levels of polarisation (polarisation fraction or degree~$>60\%$) in their sample; while \polar~and \asat~measurements find a low level of polarisation in their sample. In a few cases, \polar~\cite[GRB170114A]{Burgess2019,zhang19} and \asat/CZTI~\cite[GRB160821A]{Sharma2019} have also detected a temporal evolution of polarisation fraction, asserting the highly dynamic nature of the prompt emission. A detailed summary of GRB polarisation measurements can be found in~\citen{Gill2021}.

The limited number of measurements and low level of observed polarisation in many GRBs necessitates the need for more sensitive and dedicated GRB polarimeters. In the upcoming years, dedicated GRB polarimeters such as \emph{POLAR-2}~\citep{DeAngelis}, \emph{COSI}~\citep{Tomsick2022} and  \emph{LEAP}~\citep{McConnell2021} have been proposed to obtain more sensitive polarisation measurements. One such addition to this era would be the proposed high-energy transient mission \daksha~\citep{Bhalerao2022a,Bhalerao2022b}. When launched, it will be one of the the most sensitive high-energy telescope in the world that can also perform hard X-ray polarisation measurements. Here we present the expected hard X-ray polarisation sensitivity of \daksha.

This article is organised as follows: In Section~\ref{sec:daksha}, we give an overview of the \daksha{} mission. In Section~\ref{sec:CZT_pol}, we briefly explain the polarisation measurement principle \daksha{} detectors will utilise. In Section~\ref{sec:mass_model}, we give details of the mass model used to carry out simulations necessary for the sensitivity analysis described in the article. In Section~\ref{sec:temp_matching}, we describe the method that \daksha{} will use to measure hard X-ray polarisation. In Section~\ref{sec:pol_sensitivity}, we present the sensitivity results, and in Section~\ref{sec:conclusion} we summarise and conclude our analysis.

\section{The \daksha{} mission}\label{sec:daksha}
\daksha{} is a proposed Indian high-energy transient mission dedicated to studying electromagnetic counterparts of gravitational waves (EMGW) and GRBs~\citep{Bhalerao2022a}. Apart from its primary science goals, \daksha{} will also detect flares from magnetars, possible counterparts of Fast Radio Bursts (FRBs), outbursts from bright X-ray binaries and Active Galactic Nuclei (AGNs), hard X-ray emissions from the Sun and also Terrestrial Gamma-ray Flashes (TGFs). The details about the science goals of \daksha{} can be found in~\citen{Bhalerao2022b}.

\begin{figure}[ht]
    \centering
    \includegraphics[width=\textwidth]{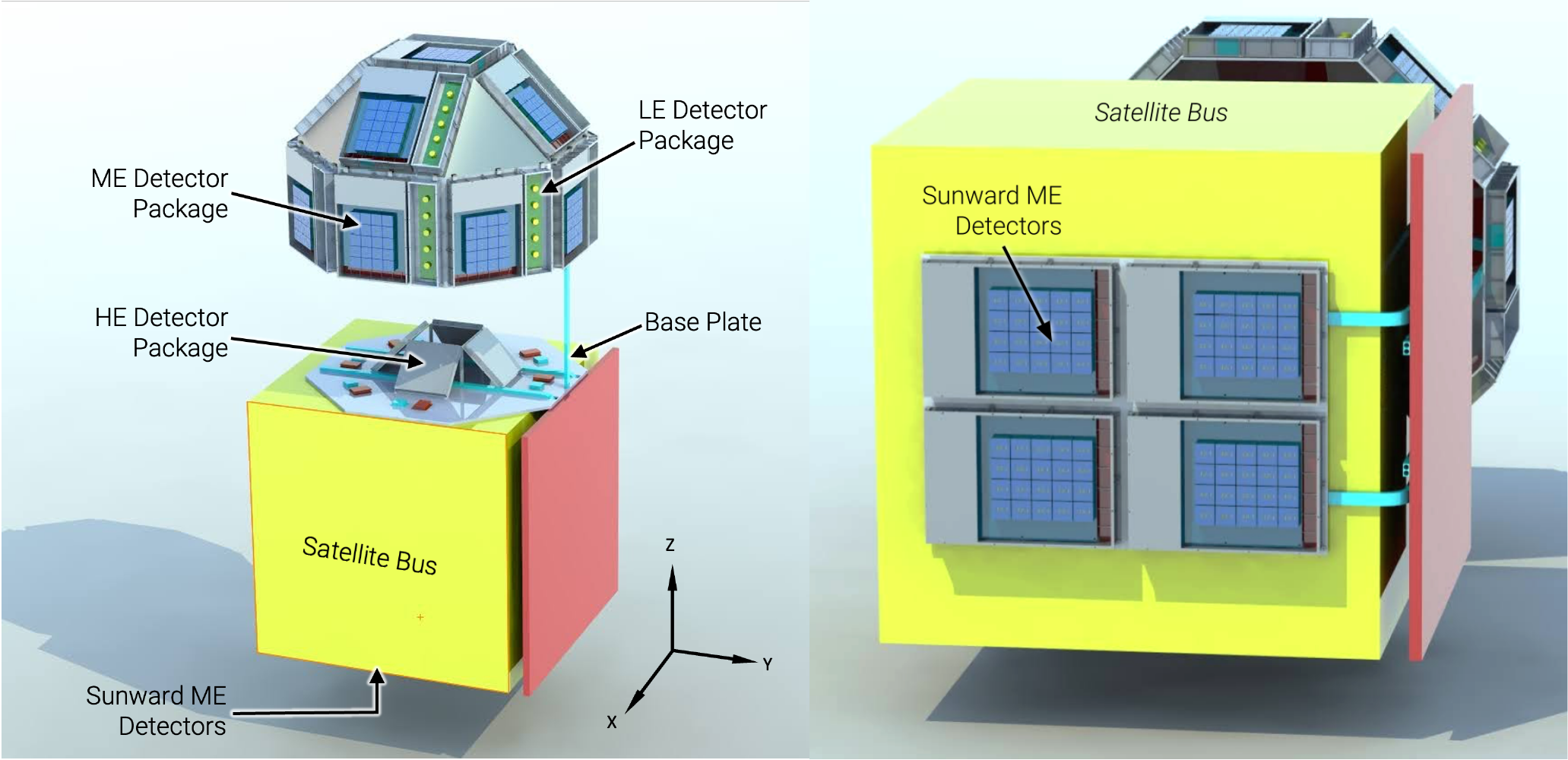}
    \caption{The design of a single \daksha{} satellite. \textbf{Left}: The payload showing 13 Low-energy (LE) and Medium-energy (ME) detector packages installed on dome-shaped frame with 13 surfaces. The four High-energy (HE) detector packages and the processing electronics are mounted inside the dome. The satellite reference frame is indicated on the bottom right. The $\theta=0\degr$ and $\phi=0\degr$ corresponds to the +ve Z-axis direction while $\theta=0\degr$ and $\phi=90\degr$ corresponds to the +ve Y-axis direction. \textbf{Right}: In addition to the 13 ME packages on the dome, four ME packages are mounted under the satellite bus. Left figure from~\citen{Bhalerao2022a}.} 
    \label{fig:daksha}
\end{figure}

\daksha{} consists of two identical satellites on opposite sides of the Earth in a Low Earth Orbit (LEO). Each satellite will cover a broad energy range (1 keV -- 1 MeV) and monitor nearly the entire sky. It will be able to detect transients onboard and send alerts over the General Coordinates Network~(GCN)\footnote{\url{https://gcn.nasa.gov/}} within a few minutes of detection. Three types of detectors are used to span the entire energy range: Silicon Drift Diodes (SDDs) form Low Energy (LE) packages to cover the 1~-- 30~keV range; Cadmium Zinc Telluride (CZT) detectors form Medium Energy (ME) packages to cover the 20~--~200~keV range; and lastly Sodium Iodide (NaI) scintillators coupled with Silicon Photo-multipliers (SiPM) cover the 100~--~1000 keV range. The current design of \daksha{} is shown in Figure~\ref{fig:daksha}, and more information about instrument details can be found in~\citen{Bhalerao2022a}.

The hemispherical arrangement of 13 ME and LE packages each gives nearly uniform coverage of half the sky. Photons from a GRB located in this half will be incident on multiple faces, all at varying angles. In the medium energy range, the sensitivity is extended to the remaining half of the sky by 4 sunward pointing ME packages.

\section{Hard X-ray polarimetry with CZT detectors}\label{sec:CZT_pol}
A single ME package onboard \daksha{} contains 20 CZT detectors in a 4$\times$5 grid with each detector having size 39.06$\times$39.06$\times$5 mm$^3$. Each detector has 256 pixels of size $\sim$2.46$\times$2.46~mm$^2$ with thickness of 5~mm. The small pixel size ($\sim2.5$ mm) and relatively large thickness (5 mm) of \daksha{} CZT detectors make them good hard X-ray polarimeters that work on the principle of Compton scattering. The polarimetric capabilities of these detectors have been discussed and demonstrated before on AstroSat/CZTI~\citep{chattopadhyay14,vadawale17,Vaishnava2022}. Here we give a brief overview of the principle for completeness.

\begin{figure}[h]
    \centering
        \includegraphics[width=0.75\textwidth]{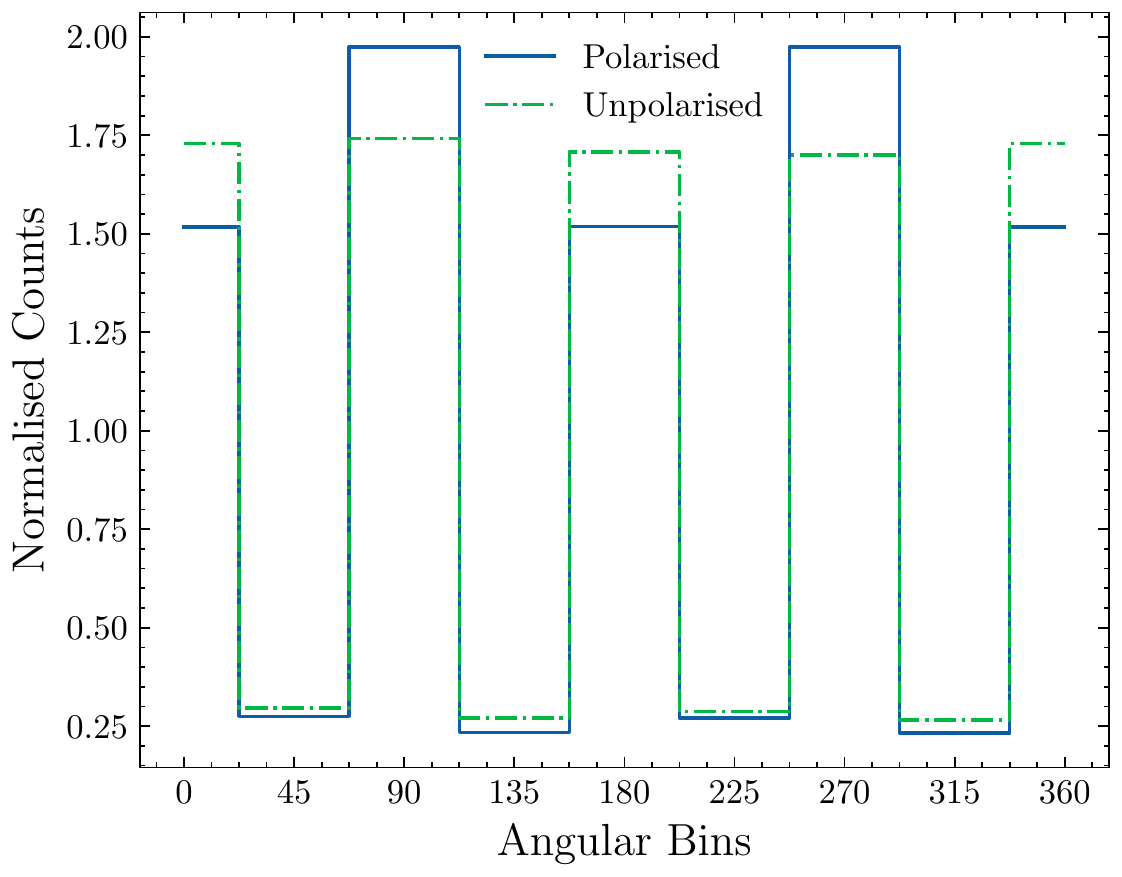}
    \caption{An example of a simulated azimuthal histograms created from the Compton event pairs detected from a modelled CZT detector. The blue curve represents the observed modulation in case of polarised photons while the green curve shows the modulation in case of unpolarised photos.}\label{fig:histogram_example}
\end{figure}

When a photon undergoes Compton scattering, it is scattered preferentially in the direction perpendicular to the direction of the polarisation vector (i.e. direction perpendicular to the electric field vector). Above the incident energy of $\sim$100 keV, Compton scattering interactions in the CZT detector can create two-pixel events where one pixel acts as a scatterer while the other acts as an absorber. The scattered photon has lower energy than the incident photon, leading to a lower mean free path within the detector. Given the pixel size, this scattered photon is likely to get absorbed in one of the adjacent pixels. A small fraction of scattered photons get absorbed in the next-neighboring pixels, considering them increases the chance coincidence probability and hence are not considered for this analysis. These ``Compton event'' pair positions can be mapped onto a square grid of 3$\times$3 pixels, with the central pixel being the scattering pixel and the surrounding pixels being the absorber pixels. For a square geometry, we get an azimuthal histogram of eight angular bins, each separated by 45\degr~(Figure~\ref{fig:histogram_example}). The modulation observed in this histogram gives us the degree of polarisation (Polarisation Fraction, PF hereafter) and the angle of polarisation (Polarisation Angle\footnote{Polarisation Angle is defined with respect to the local North increasing positively towards the East.}, PA hereafter)~\cite[for more details]{chattopadhyay14}.

Apart from Compton scattering, other effects such as chance coincidence, fluorescence and escape peak event pairs can also generate two-pixel events. Hence a selection criteria is needed to separate out possible Compton events from all the two pixel events. We employ the same criteria as described by~\citen{chattopadhyay14} to select the Compton events. This criteria was decided by running \geant{} simulations and using Compton kinematics cut on energies. The criteria, applied to both background and GRB events, are as follows:
\begin{itemize}
\item To be considered a two pixel event, the two events must have the same time stamp at the microsecond timing resolution of \daksha. 
\item The events should occur in two adjacent pixels. 
\item The pixel with lower energy deposit is assumed to be the scattering pixel ($\mathrm{E_{scat}}$) while the other pixel is assumed to be the absorber ($\mathrm{E_{abs}}$) and the energy ratio $\mathrm{E_{abs} / E_{scat}}$ should be between 1 and 6.
\item The total energy deposited in both pixels combined (i.e. $\mathrm{E_{scat} + E_{abs}}$) should be between 100 and 400 keV\footnote{The 400 keV limit comes from the upper limit of recording energy in a single pixel which is $\sim$200 keV.}.
\end{itemize}

A standard method to determine the PA and PF is by fitting a cosine function to the observed azimuthal histogram~\citep{Weisskopf2010,chattopadhyay14,chattopadhyay19,Chattopadhyay2022}. However, there is an inherent asymmetry for sources that are off-axis relative to the detector boresight (as will be the case for most GRBs observed by a single ME package), and the distribution of counts in the azimuthal bins is not strictly sinusoidal~\citep{Muleri2014}. Fitting a cosine to such a distribution leads to systematic errors in determining the PA and PF. In order to avoid the systematics that can arise from the modulation fitting, we choose the template matching approach used by~\citen{Vaishnava2022} and~\citen{zhang19} for the analysis carried out in this article (see Section~\ref{sec:temp_matching} for more details). The templates are generated such that they span entire PA -- PF parameter space and are fitted with the observed data to measure the source PA and PF. To generate such templates, we undertake extensive simulations using a detailed chemical and geometrical model (called ``mass model'') of the instrument, in this case, the ME packages and the \daksha{} payload. The following section describes the mass model used for our analysis. The details of the template matching method and sensitivity analysis are described in Section~\ref{sec:temp_matching} and Section~\ref{sec:pol_sensitivity} respectively.

\section{\daksha{} mass model}\label{sec:mass_model}
We use the popular Monte-Carlo particle-matter interaction simulation toolkit \geant~\citep{Agostinelli2003} to perform \daksha{} mass model simulations. This section briefly details the prototype mass model and its implementation in \geant. Some refinements will be made in this mass model when the final payload design is frozen, but the current model includes all the critical components that will affect the polarization measurements. Before giving more details, we first define some of the \geant{} terminologies that are needed to describe the mass model. In the discussion below, the term \texttt{event} refers to a simulation of a single photon starting with its creation and its propagation through the geometry until it (and all the other secondary particles generated by it) is stopped or leaves the simulation volume. A \texttt{run} refers to a collection of all \texttt{events} that have the same input properties, i.e. photons drawn from an identical spectral (mono-energy or Band function etc.), angular and spatial distribution.

\begin{figure}
 \centering
 \includegraphics[width=\textwidth]{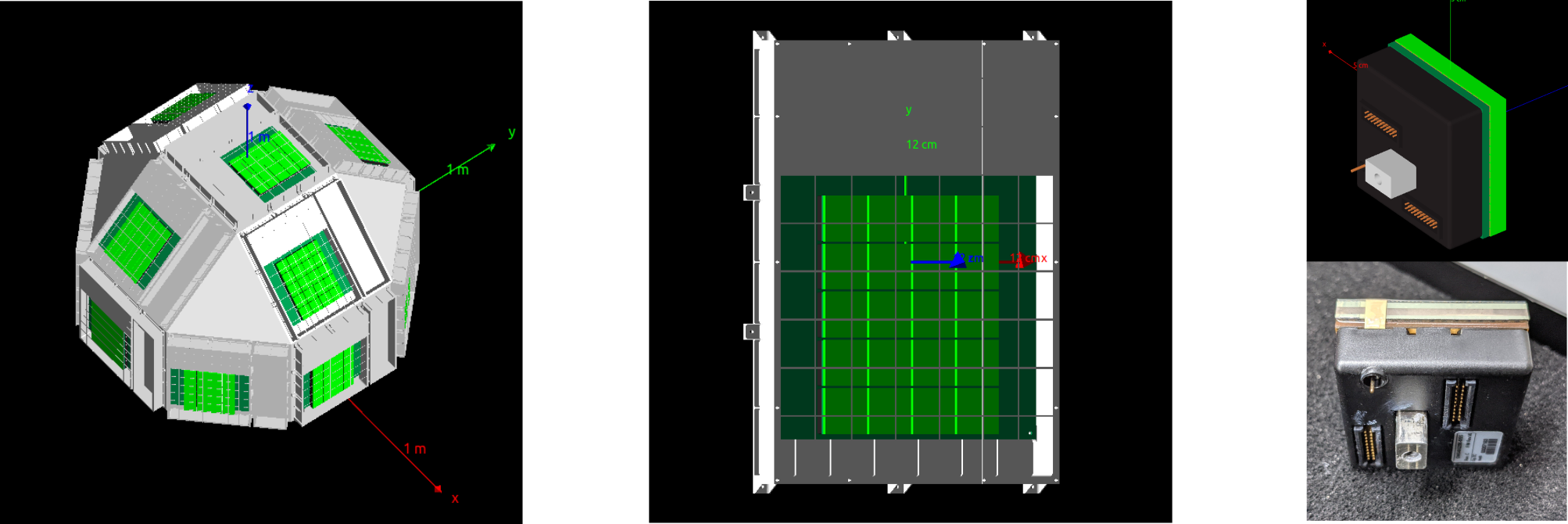}
 \caption{\geant{} rendering of \daksha{} mass model and its components. \textbf{Left}: The complete prototype model with 13 ME packages, 13 LE packages without the active volume, and the support dome. The 4 HE packages are not visible here as they are inside the dome. \textbf{Middle}: A single ME package with 20 CZT detector modules (light green squares), the front end PCB (dark green), and the support box with a top lid (grey). \textbf{Right}: A single CZT detector module rendering (top) and a photograph of a real detector module (bottom) for comparison to highlight the accuracy of modelling.}\label{fig:dakshaMM}
\end{figure}

\subsection{Geometry}
Figure~\ref{fig:dakshaMM} shows the \geant{} rendering of the prototype \daksha{} mass model and some of its sub-components. The current version consists of 13 ME packages and 13 LE packages (without the SDD detector volume) in the current configuration, four HE packages in the current configuration, and the support dome. The satellite bus has not been finalised and hence is not included in this simulation. To ensure that this does not affect our results, we are only considering GRBs in the top hemisphere.

The most critical component for our analysis, the single CZT detector module, has been modelled to a mass accuracy of better than 5\%, and the model includes most of the internal components (e.g., the Application Specific Integrated Circuit (ASIC) board, heat sink, front end Printed Circuit Board; PCB). For PCBs, individual components are not modelled and they are modelled as sheet of equivalent mass and a typical PCB composition~\citep{Hanafi2012}. For the HE package, only the NaI active volume and the Aluminium shield around the volume have been modelled. Other components like the SiPMs do not have enough material to significantly affect our simulations, but would add computational burden, hence they have been excluded.

\subsection{Physics and Tracking}
The simulation currently only tracks electromagnetic interactions between 250~eV to 100~GeV using the built-in \texttt{G4LivermorePolarisedPhysics} physics list. Atomic de-excitation is activated to produce secondary particles via fluorescence and Auger electron escape.

We have used the sensitive detector approach of \geant{} to track the interactions in the active volume. The CZT volume has been divided into 256 pixels, and each pixel acts as an individual sensitive detector\footnote{Note that although individual pixel is a sensitive detector, there is no gap between pixels. They are modelled as a parameterised volume filling the entire logical volume equivalent to a single CZT crystal of dimension 39.06 $\times$ 39.06 $\times$ 5 mm$^3$.}. All the electrons produced by the primary (or secondary) photon are tracked along with their energy deposits until they are stopped. If these electrons deposit non-zero energy in a pixel volume, the energy deposit is accumulated and the total energy deposit in that pixel is returned at the end of every \texttt{event}.

\subsection{Input and output}
The \texttt{G4GeneralParticleSource} class is used to generate input photons. The input characteristics of photons (e.g. spectral, positional, and angular distribution) are defined using the standard \geant{} input \texttt{macro} files. Currently, a planar or isotropic source with either mono energy or custom input spectra are used depending upon the simulation configuration (see Sections~\ref{sec:temp_hist}, ~\ref{sec:avg_GRB_hist} and~\ref{sec:avg_bkg_hist}).

The output of a single \texttt{run} is a FITS file similar to the ``time-tagged event files (TTE)'' produced by many X/gamma-ray instruments operating in the event mode. The file stores the event ID of the detected event (this is equivalent to event time that is stored in case of the real data), the deposited energy, and the pixel in which the event is detected\footnote{This is stored as three values: the pixel ID relative to single CZT detector module, the module ID relative to a single ME package and the ME package ID relative to the satellite frame.}. In case of multiple events, all the pixels with non-zero energy deposits are recorded individually with the same eventID\footnote{For the simulated data, to select Compton events, we use the events with same eventID in place of same timestamp.}. Along with the event info, all the metadata related to the \texttt{run} is stored in the header of the FITS extension. This includes input source geometry, the spectral and angular distribution of photons, and the input seeds to allow full reproducibility of any simulation.

\section{Polarimetry using template matching}\label{sec:temp_matching}
As mentioned in Section~\ref{sec:CZT_pol}, we use the template matching method to measure the PA and PF from the observed azimuthal histogram. The method compares (using $\chi^2$ statistics) the observed azimuthal histogram with a library of pre-computed template histograms corresponding to different PA and PF values. The following two subsections describe the steps involved in creating the library of templates and the basic principle of the template matching method.

\subsection{Template azimuthal histogram generation}\label{sec:temp_hist}
The template matching method requires a template bank that spans the entire PA/PF parameter space. The template bank can be created by simulating 100\% polarised histograms at different PAs to cover the PA space and combining them with unpolarised histograms to cover the PF space. We generate the template histogram library by running a set of \geant{} simulations for a given direction in the sky. 

For a single direction, 57 energies between 100 keV to 1 MeV are chosen with increasing step size at higher energies (typically the bin size is $\sim$ energy resolution). The upper limit on the simulation energy is decided by doing a convergence analysis such that for a typical Band spectrum, the Poisson uncertainty in 100 --- 400 keV is $\lesssim$1\%. For each energy, 18 simulations with 100\% polarised input photons are carried out by varying the polarisation angle in steps of 10\degr~from 0\degr~to 170\degr. By carrying out a limited number of simulations with finer PA spacing, we found that the number of photons in each of the eight azimuthal bins varies smoothly with PA, and the variation can be modelled as the sum of a sinusoid and its second harmonic. Thus, we can carry out simulations with 10\degr~spacing in PA, and interpolate to calculate the histograms for intermediate angles. Additional simulations with the same energies are carried out with unpolarised input photons from the same incidence direction. In total, for a given source direction, we run $57\times(18+1) = 1083$ simulations.

For each simulation, we shine 5$\times$10$^6$ photons on the entire mass model. The number is determined by carrying out a convergence estimation such that for the final template the relative error in each azimuthal bin is less than 1\%. From these simulations, azimuthal histograms at each energy and polarisation angle are extracted using the criteria explained in Section~\ref{sec:CZT_pol}.

For a source (GRB in our case) with incident spectra $S(E)$, where $E$ is the photon energy, the azimuthal histograms at each PA are co-added in the energy space by weighting them with $S(E)$. This gives us the 100\% polarised azimuthal histograms for the incident source spectra from a given direction. The same spectral scaling is applied for the unpolarised simulations to get the unpolaried histogram for the incident source spectrum. 

\begin{figure}[ht]
    \centering
     \includegraphics[width=0.9\textwidth]{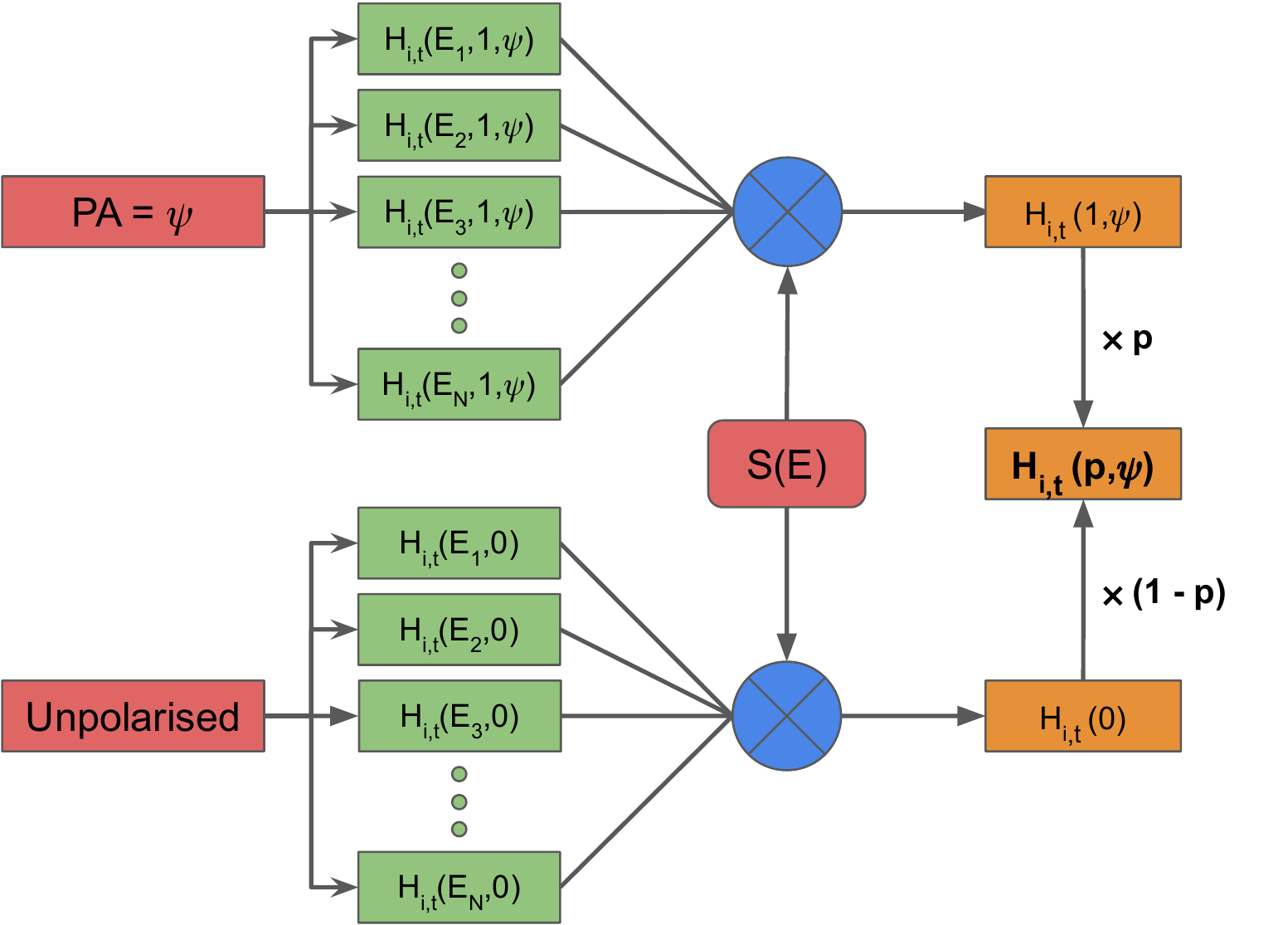}
    \caption{A schematic diagram showing creation of template histogram $H_{i, t}(p, \psi)$ for PA=$\psi$ and PF=$p$. The $H_{i, t}(E,1,\psi)$ represents the 100\% polarised template histogram at energy $E$ and PA=$\psi$, $H_{i, t}(E,0)$ represents the unpolarised template histogram at energy $E$. The $S(E)$ represents the GRB spectra. The process is repeated for different PAs and PFs to generate the template bank over the entire PA/PF parameter space.}
    \label{fig:template_gen}
\end{figure}

To create an azimuthal histogram template bank in PA/PF space for the given source, the 100\% polarised azimuthal histograms are first interpolated to a 1\degr~grid in PA space, and then they are combined with the unpolarised histograms to generate template azimuthal histograms in PF space (in PF range 0 to 1 in steps of 0.01) using the following relation:
\begin{equation} 
 H_{i, t} (p, \psi) = p \cdot H_{i, t} (1, \psi) + (1 - p) \cdot H_{i, t}(0); \hspace{10pt} 0 \leq i < 8,
\end{equation}
where $H_{i, t} (p, \psi)$ is the template azimuthal histogram for PA=$\psi$ and PF=$p$ grid point ($i$ represents the $i^{\mathrm{th}}$ azimuthal bin and $t$ stands for template histogram), $H_{i, t}(1, \psi)$ is the 100\% polarised azimuthal histogram with PA=$\psi$ and $H_{i, t}(0)$ is the unpolarised histogram. A schematic representation of template creation process is shown in the Figure~\ref{fig:template_gen}.

\subsection{Template matching by $\chi^2$ minimisation}\label{sec:chi_sq_metric}
We use $\chi^2$ statistics to quantitatively compare the observed azimuthal histogram with the library of template azimuthal histograms.  We define a $\chi^2(p, \psi)$ as:
\begin{equation}\label{eqn:chi_sq}
\chi^2(p, \psi) = \displaystyle\sum_{i=0}^{i=7}\frac{\left[\bar{H}_{i,o}(p,\psi)^{} - \bar{H}_{i, t}(p,\psi)\right]^2}{\bar{\sigma}_{i, o}^2(p,\psi) + \bar{\sigma}_{i,t}^2(p,\psi)}
\end{equation}
Where $\bar{H}_{i,o}(p,\psi)$ and $\bar{H}_{i, t}(p,\psi)$ are normalised observed and template histograms for PF value of $p$ and PA value of $\psi$ respectively and $\bar{\sigma}_{i, o}^2(p,\psi)$ and $\bar{\sigma}_{i,t}^2(p,\psi)$ are errors on them. The normalisation is done by dividing by total number of counts in all eight bins, i.e. by $N = \sum H_i$. We simulate our template bank such that $\bar{\sigma}_{i,t}^2(p,\psi) < 0.01 $ and hence in actual estimation of the $\chi^2(p, \psi)$ we neglect the $\bar{\sigma}_{i,t}^2(p,\psi)$ term. The $\chi^2(p, \psi)$ is estimated individually for each ME package on all 13 faces of \daksha. As observed counts on each face are independent, we add the $\chi^2 (p,\psi)$ values from different faces to estimate the total $\chi^2_{tot}(p, \psi)$:
\begin{equation}\label{eqn:chi_sq_tot}
\chi^2_{tot}(p, \psi) = \displaystyle\sum_j \chi^2_j(p, \psi)  
\end{equation}
where the sum is evaluated over the face ID $j$. The best-fit values for $p$ and $\psi$ for a given GRB are obtained by minimising the  $\chi^2_{tot}(p, \psi)$ over the predetermined grid of PA/PF values. The errors on the best-fit parameters are obtained using the confidence contours for a two-parameter $\chi^2$ distribution with $\Delta\chi^2_{tot}(p, \psi)$ = 2.3, 6.18, and 11.83 corresponding to the confidence levels of 1$\sigma$, 2$\sigma$ and 3$\sigma$ respectively with $\Delta\chi^2_{tot}(p, \psi)$ defined as:
\begin{equation} \label{eqn:delta_chi_sq}
\Delta\chi^2_{tot}(p, \psi) = \chi^2_{tot}(p, \psi) - \mathrm{min}(\chi^2_{tot}(p, \psi))
\end{equation}
For the minimisation process, we only use the top five faces (arranged in a decreasing order of effective area for a given direction) to compute the $\chi^2_{tot}(p, \psi)$ as our tests show that adding more faces does not have a significant impact on the results. Hence, for all the analyses described below, only the relevant five faces are used in the template-matching process.

\section{Polarimetric sensitivity of \daksha}\label{sec:pol_sensitivity}
The most common way to express the sensitivity of a polarimeter is in terms of the ``Minimum Detectable Polarisation" (MDP) that can be achieved with it. Formally, MDP is defined as: \textsl{``The degree of polarization corresponding to the amplitude of modulation that has only a 1\% probability of being detected by chance"}~\citep{Weisskopf2010}. It is computed as follows:
\begin{equation}\label{eqn:MDP_def}
    \mathrm{MDP} = \frac{4.29}{\mu_{100} R_s}\left[\frac{R_s + R_B}{T}\right]^{1/2}
\end{equation}
where $\mu_{100}$ is the modulation factor for 100\% polarised light\footnote{The modulation factor $\mu_{100}$ is broadly defined as the ratio of the semi-amplitude of the modulation curve to its mean value, see equation 7 in~\citen{Muleri2014} for the exact definition.}, $R_s$ is the source count rate, $R_B$ is the background count rate and $T$ is the total source exposure. However, this is strictly valid only for on-axis incidence as the equation is derived assuming ideal sinusoidal variation~\cite[for derivation]{Weisskopf2010}. As mentioned in Section~\ref{sec:temp_matching}, the observed modulation is not sinusoidal for off-axis incidence and particularly for the pixellated CZT detectors, hence the above equation is not strictly valid. Therefore to quantify the polarisation measurement sensitivity of \daksha{}, we use a Monte Carlo-based approach which stems from the basic definition of MDP following the approach presented in~\citen{Vaishnava2022}. In this section, we describe the Monte Carlo method (called \mcmdp hereafter) used to measure the MDP for \daksha{}, show the sensitivity results obtained using the MC-MDP, and verify our sensitivity by injecting two GRBs and recovering their polarisation.

\subsection{Minimum Detectable Polarisation using Monte Carlo approach}\label{sec:MC_MDP}
The definition of MDP states that if we repeatedly measure polarisation for an unpolarised source, in 99\% of the cases we should measure the PF less than the MDP. In other words, if we measure PF for a large number of realisations of azimuthal histograms from an unpolarised GRB, the 99$^\mathrm{th}$ percentile of the measured PF distribution gives the value of MDP for a given source direction and spectrum. The \mcmdp uses this definition to measure MDP for a given polarimeter.

Mathematically, a single realisation of an unpolarised observed azimuthal histogram corresponding to a GRB with fixed spectral parameters and duration $T_{grb}$ can be represented as:
\begin{equation}\label{eqn:hist_sampling}
    H_{i,o}(0) = P(\overline{H}^{grb}_i  T_{grb}) + P(\overline{H}^{bkg}_i T_{grb}) \\ - P(\overline{H}^{bkg}_i  T_{bkg}) \frac{T_{grb}}{T_{bkg}}
\end{equation}
where $P(\lambda)$ represents Poisson random variable with a mean $\lambda$, $\overline{H}^{grb}_i$ is the mean (time-averaged, in terms of counts/s)  azimuthal histogram (without any background) for an unpolarised GRB of a given fluence, $T_{grb}$ is the GRB duration, $\overline{H}^{bkg}_i$ is the mean (time-averaged, in terms of counts/s) background azimuthal histogram and $T_{bkg}$ is the background duration. Using this equation, one can Poisson sample large number of azimuthal histograms and measure the PA/PF value for each realisation. The MDP is then obtained from the cumulative distribution of measured PF values. The mean histograms for GRB and background are sampled by running \geant{} simulations. The details of the \mcmdp method in case of MDP estimate for \daksha{} are explained in the next section.

\subsection{Minimum Detectable Polarisation computation for \daksha}\label{sec:MDP_daksha}
\begin{figure}[h]
    \centering
    \includegraphics[width=\textwidth]{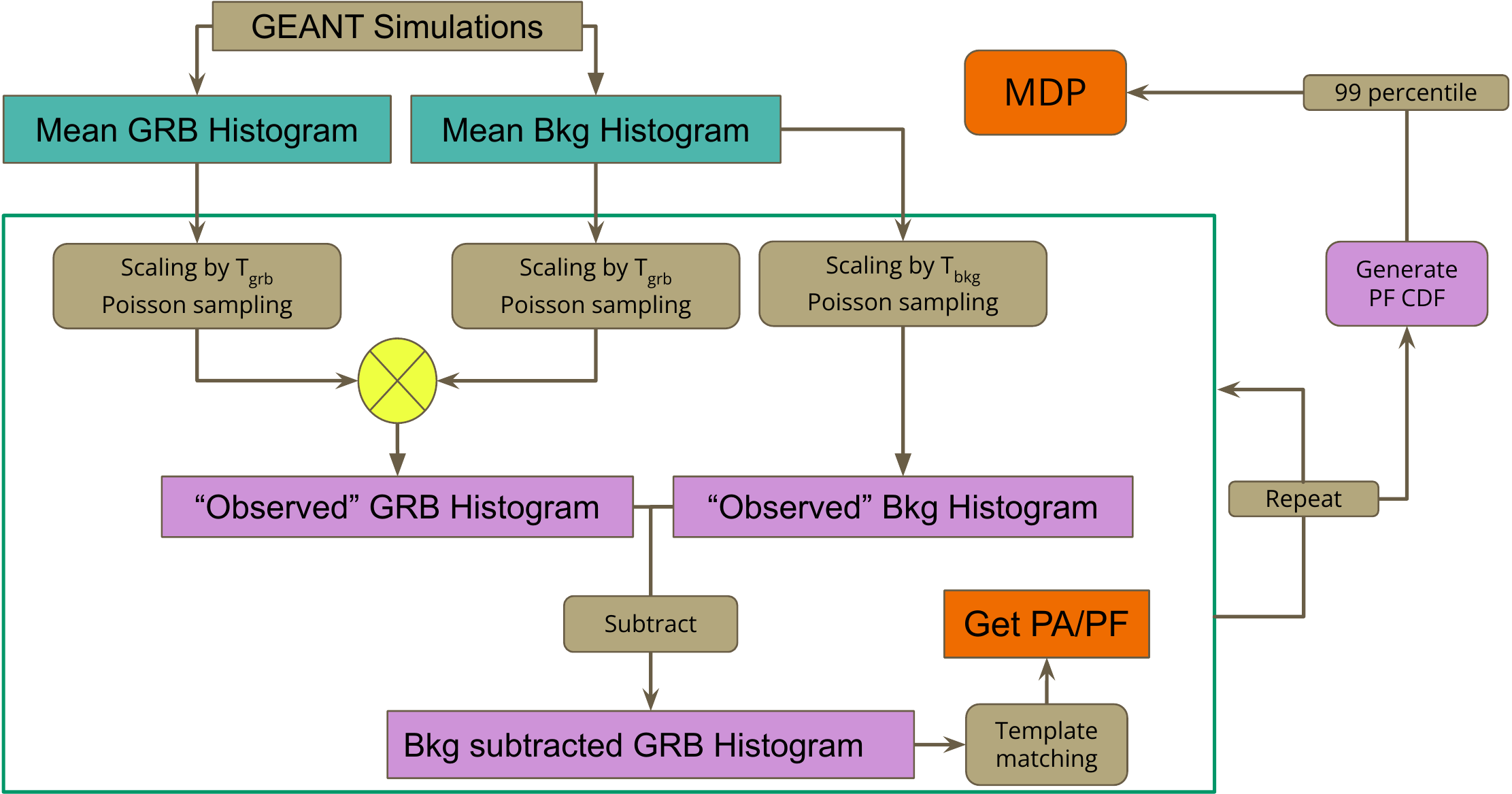}
    \caption{A diagram showing the steps involved in the \mcmdp method that is used to determine MDP for \daksha. The steps in green square are repeated 20,000 times to get the measured PF distribution from unpolarised light which is then used to determine the MDP.}\label{fig:mc_mdp_method}
\end{figure}

To estimate MDP for a given fluence and a given incident direction using \mcmdp, we create 20,000 realisations of the ``source" azimuthal histograms. Each realisation is then subjected to the template matching fitting explained in Section~\ref{sec:temp_matching} to get the measured PA and PF values. The cumulative histogram of the PF values obtained from the 20000 realisations is used to estimate the MDP. For this analysis, we fix the $T_{grb}$ to be 30~s\footnote{This value corresponds to the approximate peak of the $\mathrm{T_{90}}$ distribution in the \fermi/GBM catalogue by~\citen{Kienlin2020}.} and $T_{bkg}$ to be 1000~s. The mean GRB and background histograms are computed from \geant{} simulations as explained below. A schematic representation of the entire process is shown in Figure~\ref{fig:mc_mdp_method}.

\subsubsection{Mean GRB azimuthal histograms}\label{sec:avg_GRB_hist}
The mean GRB azimuthal histogram is created by shining a large number of \textit{unpolarised} GRB photons onto the \daksha{} mass model for a given direction. The input spectra for these simulations follow the Band~\citep{band1993batse} function with parameters: $\alpha = -1.08$, $\beta = -2.14$, E$_{\mathrm{peak}}$ = 196 keV and norm~=~1 ph/cm$^2$/s/keV at 100 keV\footnote{The parameters are taken from the \fermi/GBM catalogue by Gruber et. al. 2014~\citen{gruber2014fermi}. The values indicated as  ``BEST"  in Table 4 have been chosen.}. We simulate a total of 65,465,240\footnote{The number is obtained by integrating the Band function with the given parameters over circular planar source by assuming a GRB duration of 100~s. Note that the duration here corresponds to the simulation carried out to determine the ``mean'' GRB azimuthal histogram and not the GRB duration considered for estimating the MDP.} input photons in the 90~--~1000 keV energy range. The input energy range is fixed to 90~--~1000 keV as we are only interested in the 100 -- 400 keV range for the polarisation measurements. From this simulation, we extract the azimuthal histograms as explained in the Section~\ref{sec:CZT_pol} for each of the top five faces of \daksha. These histograms are then scaled appropriately to compute the mean GRB histogram $\overline{H}^{grb}_i$.

\subsubsection{Mean background azimuthal histograms}\label{sec:avg_bkg_hist}
The mean background azimuthal histogram is created by shining large number of photons onto the \daksha{} mass model from the inner surface of a model spherical source of radius of 5~m with a cosine-biased angular distribution to ensure an isotropic flux. The emission angle of photons is restricted between 0\degr~and 6\degr~from the surface normal to the sphere to avoid shinning input photons that do not reach the mass model~\cite[see section 4.2 for details]{Campana2013}. We only consider photon background components, i.e. Cosmic X-ray Background (CXB), reflection of CXB from Earth's atmosphere (reflection hereafter) and hard X-ray albedo of Earth (albedo hereafter) as our preliminary study shows that these components are dominant. As CXB and reflection plus albedo have different directional origin, we carry out two different simulations, one with only CXB spectrum as input and one with reflection plus albedo spectrum as input. We assume that \daksha{} is pointing away from the Earth, i.e. the top face of \daksha{} is oriented in an opposite direction to the centre of the Earth, and combine both the simulations by weighting them appropriately by expected solid angles. The spectral models for all three components are taken from~\citen{Cumani2019}. From these simulation, we extract the azimuthal histograms as explained in the Section~\ref{sec:CZT_pol} for each of the top five faces of \daksha. These histograms are then scaled appropriately to compute the mean background histogram $\overline{H}^{bkg}_i$.

\subsection{Polarisation sensitivity results for \daksha}
\begin{figure}[h]
    \centering
    \includegraphics[width=0.75\textwidth]{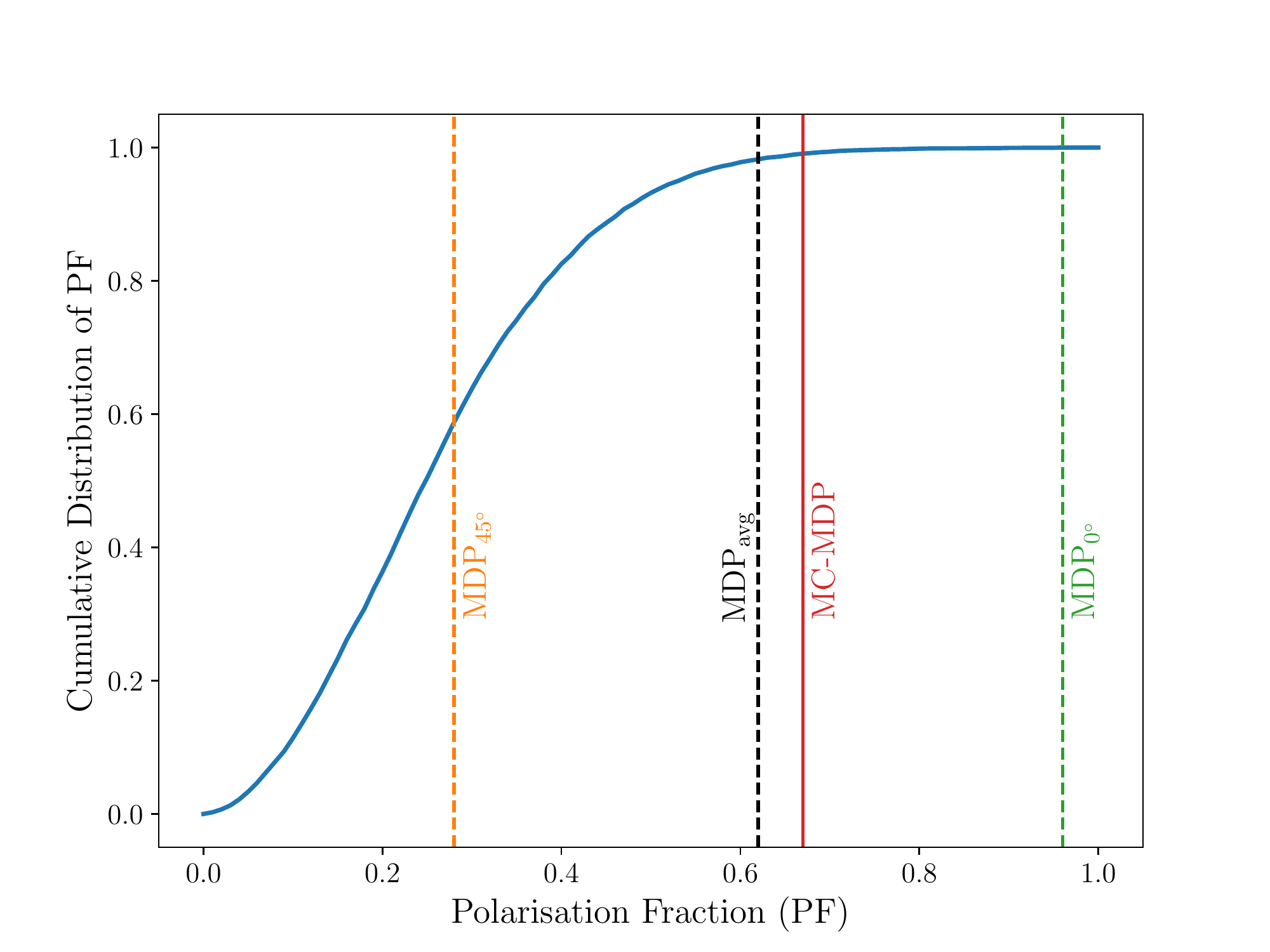}
    \caption{Comparison of MDP value obtained using the \mcmdp method with MDP value obtained using Equation~\ref{eqn:MDP_def} for a normal incidence on a single CZT detector module. The blue curve represents the cumulative distribution (CDF) of the Polarisation Fraction (PF) values obtained from the \mcmdp method. The 99$^\mathrm{th}$ percentile of of this CDF gives the MDP (red vertical line). The green and orange vertical lines show the MDP value obtain using the Equation~\ref{eqn:MDP_def} for 0\degr and 45\degr Polarisation Angle (PA), respectively. The MDP depends on PA for the analytical case because the modulation obtained from the pixellated CZT detectors is sensitive to the incident polarisation angle (see Figure 15 in ~\citen{chattopadhyay14}). From the figure it can be seen that the \mcmdp is close to the average of the two analytical values (dashed black line).}\label{fig:mdp_on_axis_test}
\end{figure}

First, to test the accuracy our method, we compare the \mcmdp with the analytical MDP (Equation~\ref{eqn:MDP_def}) for a normal incidence case. For ease of comparison, this test is performed on a single CZT detector module with incident fluence of 10$^{-3}$~\ec~in the 10 -- 1000 keV range. The comparison is shown in Figure~\ref{fig:mdp_on_axis_test}.
The reason for two MDP values in the analytical case is because $\mu_{100}$ is a function of PA in case of the pixellated CZT detector and it is minimum for PA = 0\degr~and maximum for PA = 45\degr~\cite[for more details]{chattopadhyay14}. It can be seen that the \mcmdp is close to the average of the two analytical values.
\begin{figure}[ht]
    \centering
    \includegraphics[width=\textwidth]{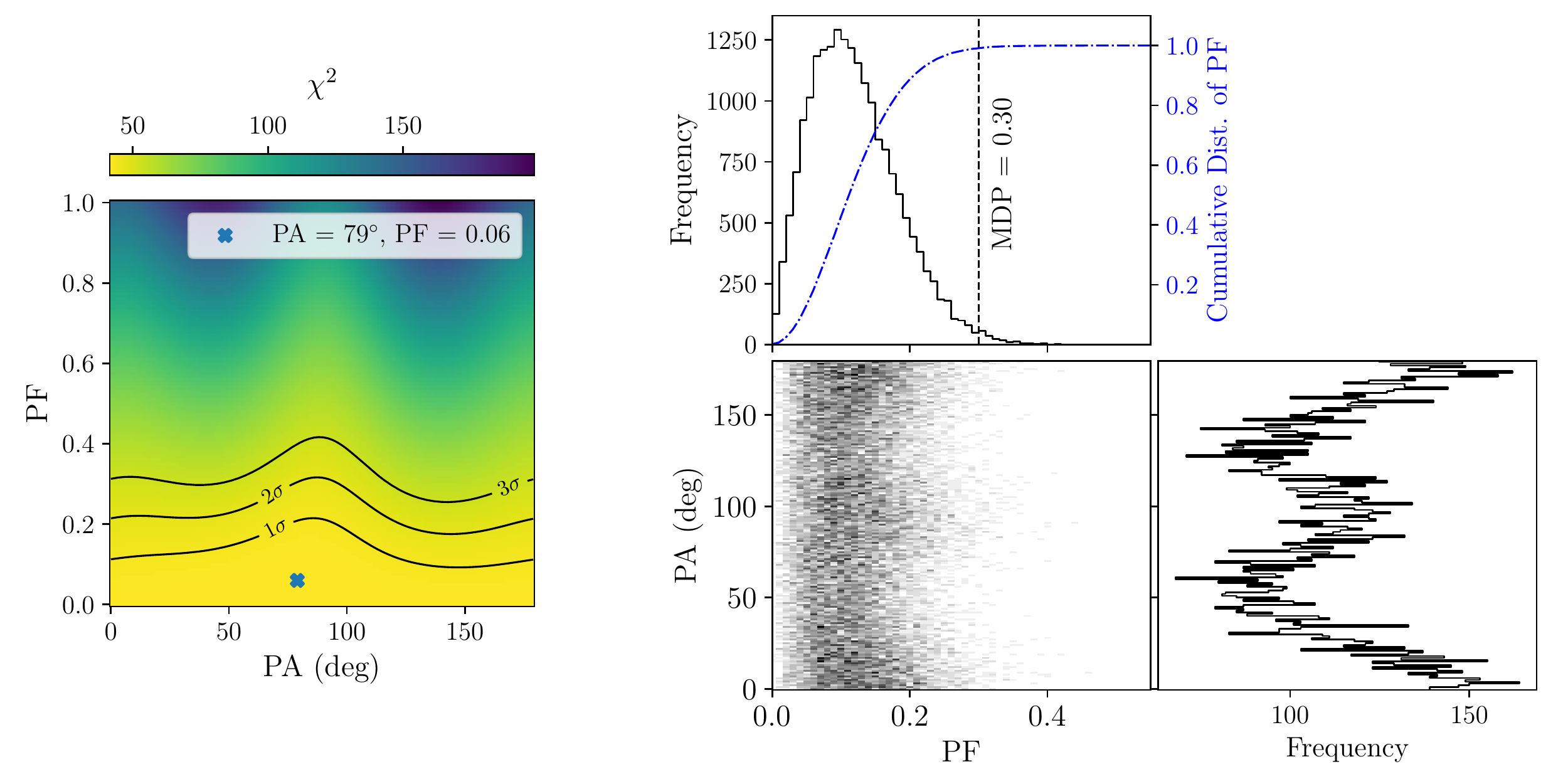}
    \caption{MDP estimation for \daksha{} ME detectors using the \mcmdp method for a near on-axis incidence ($\theta$ = 5\degr, $\phi$ = 0\degr). The incidence fluence is 10$^{-4}$~\ec~in the 10 -- 1000 keV range. {\bfseries Left}: The $\chi^2$ distribution in the PA/PF space obtained from the template matching fit for one realisation of the source azimuthal histogram. The measured value, PA~=~79\degr~and PF~=~0.06 (corresponding to the minimum $\chi^2$), is marked with the blue cross. The $\Delta  \chi^2$ contours corresponding to 1$\sigma$, 2$\sigma$ and 3$\sigma$ confidence intervals are shown with the black lines. {\bfseries Right}: Corner plot for measured PA and PF distribution for all 20,000 realisations of source azimuthal histogram. The probability and cumulative distributions of PF values (marginalised over all PAs) are shown in the top panel of the corner plot. The estimated MDP value of 0.30 is marked with a dashed line.
    }\label{fig:mdp_example}
\end{figure}
\begin{figure}
    \centering
    \includegraphics[width=0.8\textwidth]{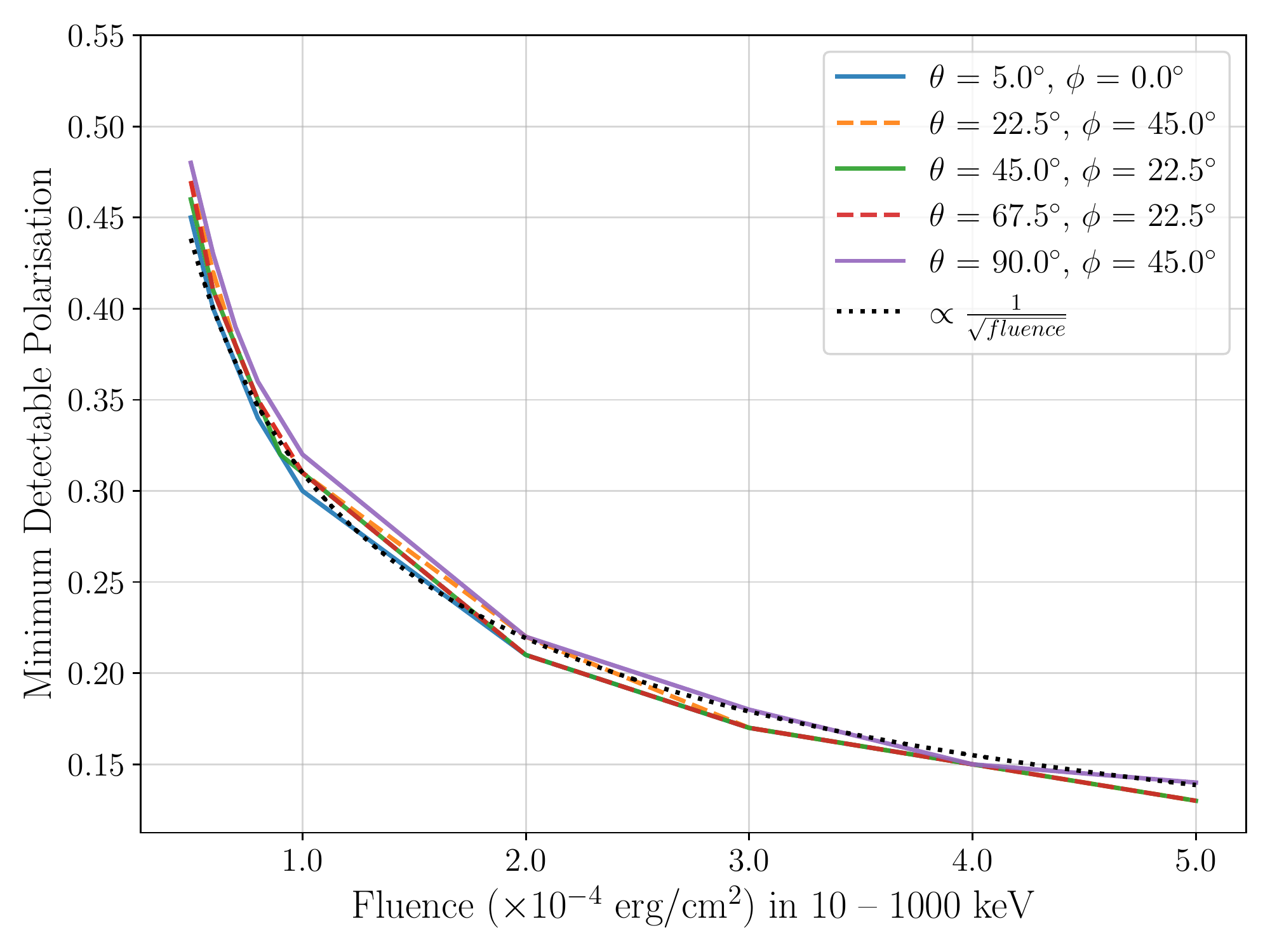}
    \caption{The variation of MDP against the incident GRB fluence for five different off-axis directions with respect to \daksha{} pointing. As illustrated by the black dashed line, the MDP drops  $\propto 1/\sqrt{\mathrm{fluence}}$ (i.e. better sensitivity) as the fluence increases. Due to the symmetry of \daksha{}, the MDP does not depend strongly on the incident direction. Only five angles are plotted here for clarity, but the other seven follow a similar trend.}\label{fig:mdp_vs_fluence}
\end{figure}

The MDP estimate for a near on-axis incidence\footnote{\daksha{} coordinate system is defined in Figure~\ref{fig:daksha} and the on-axis direction corresponds to the direction of the Z-axis.} ($\theta$ = 5\degr~and $\phi$ = 0\degr) for \daksha{} is shown in Figure~\ref{fig:mdp_example}. The left panel shows the PA/PF contour plot for one azimuthal histogram realisation, and the right panel shows the the final distribution of PA/PF for 20,000 realisations, along with the cumulative histogram of all PF values used to determine the MDP. We estimate an MDP value of 30\% with a fluence of 10$^{-4}$~\ec~in the 10 -- 1000 keV range.

The polarisation sensitivity of a polarimeter is usually correlated with source fluence, i.e., MDP decreases with an increasing GRB fluence. The sensitivity can also be a function of incident direction. To check the MDP dependence on incident direction and fluence, we estimate MDP for \daksha{} for 12 different incident directions (see Table~\ref{tab:MDP_theta_phi}) and 10 different fluence values chosen approximately logarithmically from 5$\times$10$^{-5}$~\ec~to 5$\times$10$^{-4}$~\ec. The fluences corresponds to typical bright GRBs detected by \fermi/GBM. The 12 directions are chosen in the first half of the first octant of the coordinate sphere~($\theta \in$~[0\degr, 90\degr] and $\phi \in$~[0\degr, 45\degr]) such that they span unique directions in that volume. Given the azimuthal symmetry of \daksha{}, the results obtained for these directions can be extrapolated to other $\theta, \phi$ in the top hemisphere.
\begin{table}[h]
    \begin{center}
    \caption{Incident directions (in polar coordinates) for which MDP is computed.}\label{tab:MDP_theta_phi}
    \begin{tabular}{cc}
    \hline
    $\theta$ & $\phi$ \\
    \hline
    0\degr &  5\degr \\[5pt]
    22.5\degr & 0\degr, 45\degr\\[5pt]
    45\degr & 0\degr, 22.5\degr, 45\degr\\[5pt]
    67.5\degr & 0\degr, 22.5\degr, 45\degr\\[5pt]
    90\degr & 0\degr, 22.5\degr, 45\degr\\[5pt]
    \hline
    \end{tabular}
    \end{center}
\end{table}

The Figure~\ref{fig:mdp_vs_fluence} shows the MDP variation against the incident GRB fluence for five different off-axis directions. It shows that there is no strong dependence of incident direction on MDP thanks to the symmetrical design of \daksha{} ME packages. The dependence of MDP on fluence shows an expected behaviour, and the MDP drops $\sim 1/\sqrt{\mathrm{fluence}}$ as the fluence increases.

\subsection{Validation of polarization measurements}
We verify the estimated polarisation sensitivity for \daksha{} by injecting two GRBs detected by AstroSat/CZTI. The GRBs are chosen such that they have secure measurement of polarisation fraction from current instruments, are above the MDP and fluence threshold of \daksha{}, and would be detectable (considering orbital parameters) assuming \daksha{} was active at the time of these events. The spectral and polarimetric parameters for these GRBs are taken from~\citen{Chattopadhyay2022}. To compute the position of these GRBs in the \daksha{} frame ($\theta$, $\phi$ in polar coordinates), we use satellite orbit simulations as per the mission profile and convert the celestial coordinates of the GRB into \daksha{} frame. The spectral and polarimetric parameters of the GRB as well as the calculated $\theta$, $\phi$ are given in Table~\ref{tab:injected_GRBs}.
\begin{table}[h]
    \begin{center}
    \caption{Injected GRB parameters for the polarisation sensitivity verification.}\label{tab:injected_GRBs}
    \begin{tabular}{|c|c|c|c|c|c|c|c|c|}
    \hline
    Name & $\theta$, $\phi$ & $\alpha$ & $\beta$ & E$_\mathrm{p}$ & $\Delta$T & Fluence & PF & PA\\[3pt]
     & (deg) &  &  & (keV) &  (s)  & (erg cm$^{-2}$) & & (deg)\\[3pt]
    \hline
    GRB~180103A & 90.44, 239.67 & -1.31 & -2.24 & 273 & 165.83 & $2.3\times10^{-4}$ & 0.71 & 122.13 \\
    GRB~180914B & 34.16, 124.08 & -0.75 & -2.10 & 453 & 160.04 & $5.99\times10^{-4}$ & 0.48 & 68.41\\
    \hline
    \end{tabular}
\end{center} 
\end{table}

\begin{figure}
    \centering
        \includegraphics[width=\textwidth]{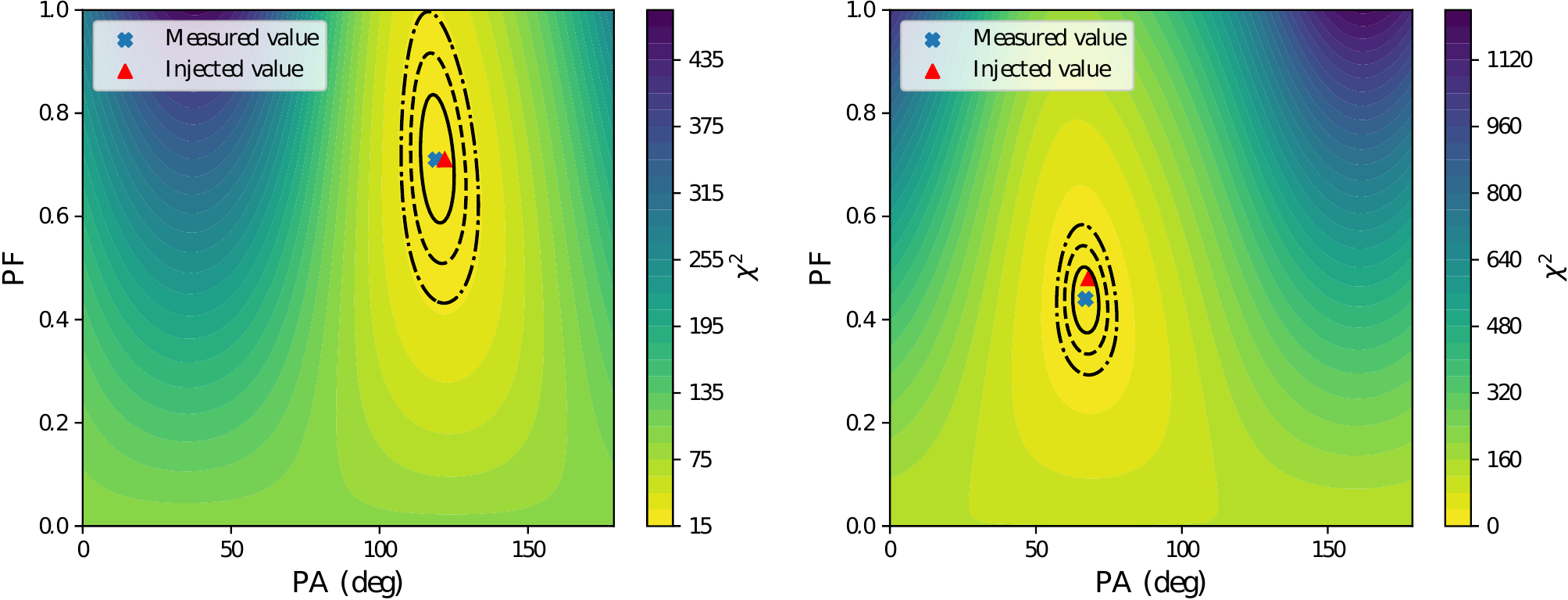}
    \caption{Figure showing measured values of PA and PF for two injected GRBs (GRB~180103A on left and GRB~180914B on right). The solid, dashed and dotted dashed black lines show the 1$\sigma$, 2$\sigma$ and 3$\sigma$ confidence intervals respectively. The injected parameters are given in table~\ref{tab:injected_GRBs}. For GRB~180103A, the measured PA/PF values are (119\degr$\pm$6\degr, 0.71$\pm$0.12) and for GRB~180914B, the measured PA/PF values are (67\degr$\pm$4\degr, 0.44$\pm$0.06). \daksha{} can recover the PA and PF accurately for both the GRBs.}\label{fig:grb_injection}
\end{figure}

The results of this test are shown in Figure~\ref{fig:grb_injection}. It can be seen that, for both GRBs, the measured values of PA and PF are close to the injected values and lie well within the 1-$\sigma$ confidence interval. In both cases, the injected PF is above the MDP, indicating \daksha{} can measure the polarisation properties of a GRB accurately, provided the PF is above the MDP threshold.

Photons for any given GRB have different angles of incidence on each ME package of the payload. This provides us with an excellent opportunity to control any systematic effects that may arise from unknown angle-dependent effects, which will likely average out over the multiple packages. For the brightest bursts, we will also be able to measure the PA and PF from individual ME packages and compare the consistency of the results.

\subsection{GRB polarisation measurement rates}
\daksha{} is expected to measure polarisation of about five GRBs per year above the fluence of $10^{-4}$~\ec\ if GRBs are highly polarised (PF $>$ 30\%). To estimate this rate, we find the number of detected GRBs above the fluence of $10^{-4}$~\ec\ by \fermi/GBM in its 14 years of operation. We scale this number by the ratio between the duty cycle of two \daksha{} satellites and \fermi/GBM and divide it by 14 to obtain the per year rate. Comparing this with \asat/CZTI, \daksha{} will be five times more sensitive. Note that although \asat/CZTI has reported 20 GRB polarisation measurements in five years, only five of them have secured polarisation measurements (other 15 are upper limits), giving a polarisation measurement rate of one per year~\citep{Chattopadhyay2022}.

\section{Conclusions}\label{sec:conclusion}

The proposed \daksha{} mission has high sensitivity for detecting Gamma Ray Bursts and other high energy transients. We can exploit the creation of two-pixel events by Compton scattering of incident photons to measure the polarisation of the source in the 100 -- 400~keV range. This ability has already been demonstrated on \asat/CZTI, and \daksha{} is poised to surpass CZTI's polarisation sensitivity.

In this article, we have discussed the method that will be used to measure hard X-ray polarisation using \daksha{} CZT detectors. The method uses a template matching approach where pre-computed templates (over a grid of PA/PF values) are compared with the observed modulation to measure the polarisation angle (PA) and polarisation fraction (PF) of the source. We have performed detailed \geant{} simulations using the mass model of \daksha{} to quantify the polarisation measurement sensitivity of \daksha. The sensitivity has been quoted using the standard Minimum Detectable Polarisation (MDP) metric. To account for the non-normal incidence directions of photons, we have adopted a new Monte-Carlo-based approach to compute the MDP. This method gives consistent results with the analytical formula for on-axis cases, but can be readily generalised to other angles.

Our results show that thanks to the symmetrical design of \daksha, for a given fluence, MDP does not depend strongly on the incident direction (with respect to the satellite pointing), and hence \daksha{} will have a near-uniform polarisation measurement sensitivity for half of the sky. For a fluence of $10^{-4}$~\ec\ in the energy range 10 -- 1000 keV, we obtain MDP of 30\% for \daksha. Given this sensitivity, we predict that if GRBs are highly polarised, \daksha{} can confidently measure polarisation for at least five GRBs per year; five times better than \asat/CZTI. \daksha{} is likely to be operational during the same period as other dedicated GRB polarisation missions such as \emph{POLAR-2}, \emph{COSI} and \emph{LEAP}. Given the detection sensitivity and all-sky coverage of \daksha{}, it will detect many GRBs simultaneously with these missions, and joint analysis of such GRBs will play an important role in understanding the prompt emission. A more detailed analysis of GRB polarisation measurement statistics with \daksha{} and its implication on breaking the degeneracy in the proposed physical models for prompt emission will be carried out in subsequent works. Overall, when launched, \daksha{} will play an important role in the field of GRB polarisation.

\section*{Acknowledgments}
We thank the Space Program Office (SPO) of the Indian Space Research Organisation for its Announcement of Opportunity for space astrophysics missions, under which \daksha\ was proposed. Development of the \daksha\ Medium Energy Package laboratory model was started with funding support from SPO, and continued with support from all partner organisations. We thank the administrative and support staff at all partner institutes for their help in all \daksha-related matters.

We want to thank Dr. Shabnam Iyyani from IISER Thiruvananthapuram, and Dr. Tanmoy Chattopadhyay from Kavli Institute of Particle Astrophysics and Cosmology, Stanford University for their valuable inputs, comments and discussions that have helped improve the manuscript. S.M. would like to thank Dr. Ajay Vibhute and Mr. Dhanraj Borgaonkar of IUCAA, Pune for their help to configure the High Performance Computing cluster at IUCAA which was essential for the simulations performed in this article. S.P.T. is a CIFAR Azrieli Global Scholar in the Gravity and Extreme Universe Program and this work was supported from the CIFAR Azrieli Fellowship Grant.

\section*{Softwares}
Numpy~\citep{numpy}, Scipy~\citep{2020SciPy-NMeth}, Matplotlib~\citep{matplotlib},
Astropy~\cite[\url{http://www.astropy.org}]{2013A&A...558A..33A,2018AJ....156..123A}, \\
Ephem~ [\url{https://pypi.python.org/pypi/pyephem/}], GEANT4~\cite[\url{https://geant4.web.cern.ch/}]{Agostinelli2003}

\section*{Code, Data, and Materials Availability}
The codes are in a continuous state of development and can be made available on a reasonable request to the authors.


\bibliography{daksha_pol}   

\begin{thebibliography}{10}

\bibitem{discovery}
R.~W. {Klebesadel}, I.~B. {Strong}, and R.~A. {Olson}, ``{Observations of
  Gamma-Ray Bursts of Cosmic Origin},'' {\em ApJL} {\bf 182}, L85  (1973).

\bibitem{Rees1992}
M.~J. {Rees} and P.~{Meszaros}, ``{Relativistic fireballs - Energy conversion
  and time-scales.},'' {\em \mnras} {\bf 258}, 41  (1992).

\bibitem{piran05}
T.~{Piran}, ``{The physics of gamma-ray bursts},'' {\em Reviews of Modern
  Physics} {\bf 76}, 1143--1210  (2004).

\bibitem{meszaros06}
P.~{M{\'e}sz{\'a}ros}, ``{Gamma-ray bursts},'' {\em Reports on Progress in
  Physics} {\bf 69}, 2259--2321  (2006).

\bibitem{eichler89}
D.~{Eichler}, M.~{Livio}, T.~{Piran}, {\em et~al.}, ``{Nucleosynthesis,
  neutrino bursts and gamma-rays from coalescing neutron stars},'' {\em \nat}
  {\bf 340}, 126--128  (1989).

\bibitem{narayan92}
R.~{Narayan}, B.~{Paczynski}, and T.~{Piran}, ``{Gamma-ray bursts as the death
  throes of massive binary stars},'' {\em \apjl} {\bf 395}, L83--L86  (1992).

\bibitem{Abbott_170817A}
B.~P. Abbott, R.~Abbott, T.~Abbott, {\em et~al.}, ``{Gravitational Waves and
  Gamma-Rays from a Binary Neutron Star Merger: GW170817 and GRB 170817A},''
  {\em \apjl} {\bf 848}, L13  (2017).

\bibitem{woosley93}
S.~E. {Woosley}, ``{Gamma-ray bursts from stellar mass accretion disks around
  black holes},'' {\em \apj} {\bf 405}, 273--277  (1993).

\bibitem{iwamoto98}
K.~{Iwamoto}, P.~A. {Mazzali}, K.~{Nomoto}, {\em et~al.}, ``{A hypernova model
  for the supernova associated with the {$\gamma$}-ray burst of 25 April
  1998},'' {\em \nat} {\bf 395}, 672--674  (1998).

\bibitem{macfadyen99}
A.~I. {MacFadyen} and S.~E. {Woosley}, ``{Collapsars: Gamma-Ray Bursts and
  Explosions in ``Failed Supernovae''},'' {\em \apj} {\bf 524}, 262--289
  (1999).

\bibitem{kumar15}
P.~{Kumar} and B.~{Zhang}, ``{The physics of gamma-ray bursts \& relativistic
  jets},'' {\em \physrep} {\bf 561}, 1--109  (2015).

\bibitem{zhang19}
S.-N. {Zhang}, M.~{Kole}, T.-W. {Bao}, {\em et~al.}, ``{Detailed polarization
  measurements of the prompt emission of five gamma-ray bursts},'' {\em Nature
  Astronomy} {\bf 3}, 258--264  (2019).

\bibitem{Gill_etal_2021_review}
R.~{Gill}, M.~{Kole}, and J.~{Granot}, ``{GRB Polarization: A Unique Probe of
  GRB Physics},'' {\em arXiv e-prints} , arXiv:2109.03286  (2021).

\bibitem{band1993batse}
D.~{Band}, J.~{Matteson}, L.~{Ford}, {\em et~al.}, ``{BATSE Observations of
  Gamma-Ray Burst Spectra. I. Spectral Diversity},'' {\em ApJ} {\bf 413}, 281
  (1993).

\bibitem{gruber2014fermi}
D.~Gruber, A.~Goldstein, V.~W. von Ahlefeld, {\em et~al.}, ``The fermi gbm
  gamma-ray burst spectral catalog: four years of data,'' {\em The
  Astrophysical Journal Supplement Series} {\bf 211}(1), 12  (2014).

\bibitem{iyyani15}
S.~{Iyyani}, F.~{Ryde}, B.~{Ahlgren}, {\em et~al.}, ``{Extremely narrow
  spectrum of GRB110920A: further evidence for localized, subphotospheric
  dissipation},'' {\em \mnras} {\bf 450}, 1651--1663  (2015).

\bibitem{Zhang2016}
B.-B. Zhang, Z.~L. Uhm, V.~Connaughton, {\em et~al.}, ``{SYNCHROTRON} {ORIGIN}
  {OF} {THE} {TYPICAL} {GRB} {BAND} {FUNCTION}{\textemdash}a {CASE} {STUDY}
  {OF} {GRB} 130606b,'' {\em The Astrophysical Journal} {\bf 816}, 72  (2016).

\bibitem{toma08}
K.~{Toma}, T.~{Sakamoto}, B.~{Zhang}, {\em et~al.}, ``{Statistical Properties
  of Gamma-Ray Burst Polarization},'' {\em Astrophysical Journal} {\bf 698},
  1042--1053  (2009).

\bibitem{covino16}
S.~{Covino} and D.~{Gotz}, ``{Polarization of prompt and afterglow emission of
  Gamma-Ray Bursts},'' {\em Astronomical and Astrophysical Transactions} {\bf
  29}, 205--244  (2016).

\bibitem{mcconnell16}
M.~L. {McConnell}, ``{High energy polarimetry of prompt GRB emission},'' {\em
  \nar} {\bf 76}, 1--21  (2017).

\bibitem{gill18}
R.~{Gill}, J.~{Granot}, and P.~{Kumar}, ``{Linear polarization in gamma-ray
  burst prompt emission},'' {\em \mnras} {\bf 491}, 3343--3373  (2020).

\bibitem{Kole2020}
M.~Kole, N.~{De Angelis}, F.~Berlato, {\em et~al.}, ``{The POLAR gamma-ray
  burst polarization catalog},'' {\em Astron. Astrophys.} {\bf 644}  (2020).

\bibitem{chattopadhyay19}
T.~{Chattopadhyay}, S.~V. {Vadawale}, E.~{Aarthy}, {\em et~al.}, ``{Prompt
  Emission Polarimetry of Gamma-Ray Bursts with the AstroSat CZT Imager},''
  {\em \apj} {\bf 884}, 123  (2019).

\bibitem{Chattopadhyay2022}
T.~Chattopadhyay, S.~Gupta, S.~Iyyani, {\em et~al.}, ``{Hard X-Ray Polarization
  Catalog for a Five-year Sample of Gamma-Ray Bursts Using AstroSat CZT
  Imager},'' {\em Astrophys. J.} {\bf 936}, 12  (2022).

\bibitem{Willis2005}
D.~R. Willis, E.~J. Barlow, A.~J. Bird, {\em et~al.}, ``{Evidence of
  polarisation in the prompt gamma-ray emission from GRB 930131 and GRB
  960924},'' {\em Astron. Astrophys.} {\bf 439}, 245--253  (2005).

\bibitem{Yonetoku2012}
D.~{Yonetoku}, T.~{Murakami}, S.~{Gunji}, {\em et~al.}, ``{Magnetic Structures
  in Gamma-Ray Burst Jets Probed by Gamma-Ray Polarization},'' {\em \apjl} {\bf
  758}, L1  (2012).

\bibitem{Gotz09}
D.~{G{\"o}tz}, P.~{Laurent}, F.~{Lebrun}, {\em et~al.}, ``{Variable
  Polarization Measured in the Prompt Emission of GRB 041219A Using IBIS on
  Board INTEGRAL},'' {\em Astrophysical Journal Letter} {\bf 695}, L208--L212
  (2009).

\bibitem{Gotz13}
D.~{G{\"o}tz}, S.~{Covino}, A.~{Fern{\'a}ndez-Soto}, {\em et~al.}, ``{The
  polarized gamma-ray burst GRB 061122},'' {\em MNRAS} {\bf 431}, 3550--3556
  (2013).

\bibitem{Burgess2019}
J.~M. {Burgess}, M.~{Kole}, F.~{Berlato}, {\em et~al.}, ``{Time-resolved GRB
  polarization with POLAR and GBM. Simultaneous spectral and polarization
  analysis with synchrotron emission},'' {\em \aap} {\bf 627}, A105  (2019).

\bibitem{Sharma2019}
V.~{Sharma}, S.~{Iyyani}, D.~{Bhattacharya}, {\em et~al.}, ``{Time-varying
  Polarized Gamma-Rays from GRB 160821A: Evidence for Ordered Magnetic
  Fields},'' {\em \apjl} {\bf 882}, L10  (2019).

\bibitem{Gill2021}
R.~Gill, M.~Kole, and J.~Granot, ``{GRB polarization: A unique probe of grb
  physics},'' {\em Galaxies} {\bf 9}  (2021).

\bibitem{DeAngelis}
N.~{De Angelis}, J.~M. Burgess, F.~Cadoux, {\em et~al.}, ``{Development and
  science perspectives of the POLAR-2 instrument: a large scale GRB
  polarimeter},'' in {\em Proc. Sci.},   {\bf 395}  (2022).

\bibitem{Tomsick2022}
J.~A. {Tomsick}, A.~{Lowell}, H.~{Lazar}, {\em et~al.}, ``{Soft Gamma-Ray
  Polarimetry with COSI Using Maximum Likelihood Analysis},'' in {\em Handbook
  of X-ray and Gamma-ray Astrophysics},  73  (2022).

\bibitem{McConnell2021}
M.~L. {McConnell}, M.~{Baring}, P.~{Bloser}, {\em et~al.}, ``{The LargE Area
  burst Polarimeter (LEAP) a NASA mission of opportunity for the ISS},'' in
  {\em UV, X-Ray, and Gamma-Ray Space Instrumentation for Astronomy XXII},
  O.~H. {Siegmund}, Ed., {\em Society of Photo-Optical Instrumentation
  Engineers (SPIE) Conference Series} {\bf 11821}, 118210P  (2021).

\bibitem{Bhalerao2022a}
V.~{Bhalerao}, S.~{Vadawale}, S.~{Tendulkar}, {\em et~al.}, ``{Daksha: On Alert
  for High Energy Transients},'' {\em arXiv e-prints} , arXiv:2211.12055
  (2022).

\bibitem{Bhalerao2022b}
V.~{Bhalerao}, D.~{Sawant}, A.~{Pai}, {\em et~al.}, ``{Science with the Daksha
  High Energy Transients Mission},'' {\em arXiv e-prints} , arXiv:2211.12052
  (2022).

\bibitem{chattopadhyay14}
T.~{Chattopadhyay}, S.~V. {Vadawale}, A.~R. {Rao}, {\em et~al.}, ``{Prospects
  of hard X-ray polarimetry with Astrosat-CZTI},'' {\em Experimental Astronomy}
  {\bf 37}, 555--577  (2014).

\bibitem{vadawale17}
S.~V. {Vadawale}, T.~{Chattopadhyay}, N.~P.~S. {Mithun}, {\em et~al.},
  ``{Phase-resolved X-ray polarimetry of the Crab pulsar with the AstroSat CZT
  Imager},'' {\em Nature Astronomy} {\bf 2}, 50--55  (2018).

\bibitem{Vaishnava2022}
C.~S. Vaishnava, N.~P.~S. Mithun, S.~V. Vadawale, {\em et~al.}, ``{Experimental
  verification of off-axis polarimetry with cadmium zinc telluride detectors of
  AstroSat-CZT Imager},'' {\em J. Astron. Telesc. Instruments, Syst.} {\bf 8},
  038005  (2022).

\bibitem{Weisskopf2010}
M.~C. Weisskopf, R.~F. Elsner, and S.~L. O'Dell, ``{On understanding the
  figures of merit for detection and measurement of x-ray polarization},'' {\em
  Space Telescopes and Instrumentation 2010: Ultraviolet to Gamma Ray} {\bf
  7732}, 77320E  (2010).

\bibitem{Muleri2014}
F.~Muleri, ``{On the operation of X-ray polarimeters with a large field of
  view},'' {\em Astrophys. J.} {\bf 782}, 28  (2014).

\bibitem{Agostinelli2003}
S.~Agostinelli, J.~Allison, K.~Amako, {\em et~al.}, ``{GEANT4 - A simulation
  toolkit},'' {\em Nuclear Instruments and Methods in Physics Research, Section
  A: Accelerators, Spectrometers, Detectors and Associated Equipment} {\bf
  506}, 250--303  (2003).

\bibitem{Hanafi2012}
J.~Hanafi, E.~Jobiliong, A.~Christiani, {\em et~al.}, ``Material recovery and
  characterization of pcb from electronic waste,'' {\em Procedia - Social and
  Behavioral Sciences} {\bf 57}, 331--338  (2012).
\newblock International Conference on Asia Pacific Business Innovation and
  Technology Management.

\bibitem{Kienlin2020}
A.~von Kienlin, C.~A. Meegan, W.~S. Paciesas, {\em et~al.}, ``{The Fourth
  Fermi-GBM Gamma-Ray Burst Catalog: A Decade of Data},'' {\em Astrophys. J.}
  {\bf 893}, 46  (2020).

\bibitem{Campana2013}
R.~Campana, M.~Feroci, E.~{Del Monte}, {\em et~al.}, ``{Background simulations
  for the Large Area Detector onboard LOFT},'' {\em Experimental Astronomy}
  {\bf 36}, 451--477  (2013).

\bibitem{Cumani2019}
P.~Cumani, M.~Hernanz, J.~Kiener, {\em et~al.}, ``{Background for a gamma-ray
  satellite on a low-Earth orbit},'' {\em Experimental Astronomy} {\bf 47},
  273--302  (2019).

\bibitem{numpy}
S.~van~der Walt, S.~C. Colbert, and G.~Varoquaux, ``{The NumPy Array: A
  Structure for Efficient Numerical Computation},'' {\em Computing in Science
  {\&} Engineering} {\bf 13}, 22--30  (2011).

\bibitem{2020SciPy-NMeth}
P.~Virtanen, R.~Gommers, T.~E. Oliphant, {\em et~al.}, ``{{SciPy} 1.0:
  Fundamental Algorithms for Scientific Computing in Python},'' {\em Nature
  Methods} {\bf 17}, 261--272  (2020).

\bibitem{matplotlib}
J.~D. Hunter, ``{Matplotlib: A 2D Graphics Environment},'' {\em Computing in
  Science {\&} Engineering} {\bf 9}, 90--95  (2007).

\bibitem{2013A&A...558A..33A}
{Astropy Collaboration}, T.~P. {Robitaille}, E.~J. {Tollerud}, {\em et~al.},
  ``{Astropy: A community Python package for astronomy},'' {\em \aap} {\bf
  558}, A33  (2013).

\bibitem{2018AJ....156..123A}
{Astropy Collaboration}, A.~M. {Price-Whelan}, B.~M. {Sip{\H{o}}cz}, {\em
  et~al.}, ``{The Astropy Project: Building an Open-science Project and Status
  of the v2.0 Core Package},'' {\em \aj} {\bf 156}, 123  (2018).

\end{thebibliography}
\bibliographystyle{spiejour}   



\end{spacing}
\end{document}